\documentclass[prd,twocolumn,nofootinbib]{revtex4}
\usepackage{amsmath,amssymb}
\usepackage{hyperref}
\usepackage{url}
\usepackage{graphicx}
\usepackage{multirow}
\usepackage{color}
\usepackage{mathrsfs}
\usepackage{breakurl}

\def\beq{\begin{equation}}
\def\eeq{\end{equation}}
\def\beqa{\begin{eqnarray}}
\def\eeqa{\end{eqnarray}}
\def\bit{\begin{itemize}}
\def\eit{\end{itemize}}

\newcommand{\ts}{\thinspace}

\begin{document}
\title{Three Twin Neutrinos: Evidence from LSND and MiniBooNE}
\author{
Yang Bai,$^{a}$ Ran Lu,$^{a}$ Sida Lu,$^{a}$ Jordi Salvado$^{b}$ and Ben A. Stefanek$^{a}$
\vspace{5mm}
\\
$^{a}$ \normalsize\emph{Department of Physics, University of Wisconsin-Madison,  Madison, WI 53706, USA}  \vspace{1mm} \\
$^{b}$ \normalsize\emph{Instituto de F\'isica Corpuscular, Universidad de Valencia CSIC, Valencia 46071, Spain}
}
\begin{abstract}
We construct a neutrino model of three twin neutrinos in light of the neutrino appearance excesses at LSND and MiniBooNE. The model, which includes a twin parity, naturally predicts identical lepton Yukawa structures in the Standard Model and the twin sectors. As a result, a universal mixing angle controls all three twin neutrino couplings to the Standard Model charged leptons. This mixing angle is predicted to be the ratio of the electroweak scale over the composite scale of the Higgs boson and has the right order of magnitude to fit the data. The heavy twin neutrinos decay within the experimental lengths into active neutrinos plus a long-lived Majoron and can provide a good fit, at around $4\sigma$ confidence level, to the LSND and MiniBooNE appearance data while simultaneously satisfying the disappearance constraints. For the Majorana neutrino case, the fact that neutrinos have a larger scattering cross section than anti-neutrinos provides a natural explanation for MiniBooNE's observation of a larger anti-neutrino appearance excess.
\end{abstract}
\maketitle
%%%%%%%%%%%%%%%%%%%%%%%%%%%%%%%%
\section{Introduction}
%%%%%%%%%%%%%%%%%%%%%%%%%%%%%%%
The past decade has seen a series of anomalies emerge in short baseline (SBL) neutrino oscillation experiments which cannot be explained within the three active neutrino framework of the Standard Model (SM). Here, SBL refers to experiments with the ratio of the oscillation distance over the neutrino energy, $L/E_{\nu} \sim 1 \ts \rm{m/MeV}$, which are sensitive to neutrino oscillations involving mass squared splittings $\Delta m^{2} \sim 1~\rm{eV}^{2}$. The LSND experiment \cite{Aguilar:2001ty} reports evidence of $\bar{\nu}_{\mu} \rightarrow \bar{\nu}_{e}$ oscillation consistent with $\Delta m^{2} \sim 1~\rm{eV}^{2}$, as well as a less dramatic excess for $\nu_{\mu} \rightarrow \nu_{e}$ oscillation~\cite{Athanassopoulos:1997er}. The MiniBooNE collaboration also searched for the same signal, reporting excesses in both electron and anti-electron neutrino events~\cite{Aguilar-Arevalo:2013pmq}, again suggesting oscillations of the form $\nu_{\mu} \rightarrow \nu_{e}$ and $\bar{\nu}_{\mu} \rightarrow \bar{\nu}_{e}$, consistent with the LSND results. Together, these observations lead to the tantalizing suggestion of additional ``sterile'' neutrino flavors at a mass scale of $1 \ts\rm{eV}$.

Many schemes have been considered to fit the excess, including 3 active plus $N$ sterile neutrino oscillation schemes (3+$N$), with most of the attention being focused on $N=1$ and $N=2$ \cite{Conrad:2013mka, Karagiorgi:2006jf,Giunti:2010jt,Nelson:2010hz,Kopp:2011qd,Fan:2012ca,Kuflik:2012sw,Huang:2013zga,Aguilar-Arevalo:2013pmq,Kopp:2013vaa}. While even the simple 3+1 scheme can provide a good fit to the $\nu_{\mu} \rightarrow \nu_{e}$ ($\bar{\nu}_{\mu} \rightarrow \bar{\nu}_{e}$) appearance excesses, these fits are in tension with $\nu_\mu$, $\bar{\nu}_\mu$ and $\nu_e$  disappearance constraints from MiniBooNE+SciBooNE~\cite{Cheng:2012yy,Mahn:2011ea} and LSND+KARMEN \cite{Conrad:2011ce}, respectively. To ameliorate the disappearance constraint, some authors have also considered fairly prompt decay of sterile neutrinos \cite{PalomaresRuiz:2005vf} ({\it i.e.} $m_{s}\Gamma_{s} \sim 1 \ts \rm{eV}^{2}$) to allow the decay of sterile neutrinos to active neutrinos within the experimental lengths. In most cases, a decay of this form requires the coupling of neutrinos to a new light state $\varphi$ (potentially a Majoron~\cite{Chikashige:1980qk,Chikashige:1980ui,Gelmini:1980re}), which enables the sterile neutrinos to decay through the process $\nu_{s} \rightarrow \nu_{a} + \varphi$. While some authors have also considered decays of the form $\nu_{s} \rightarrow \nu_{a} + \gamma$ \cite{Gninenko:2009ks} to explain the MiniBooNE signal, this decay cannot explain the LSND excess and we will not consider it here.

Very little attention has been focused on the 3+3 oscillation scheme, mainly in the interest of minimality and because there was no clear indication that adding a third sterile neutrino would improve the 3+2 fit. However, when the sterile neutrino sector is embedded within a model that ``mirrors" the SM particle content~\cite{Kobzarev:1966qya,Berezhiani:1995yi,FOOT199167}, the 3+3 scenario becomes natural to consider. One well motivated model of this type is the ``Twin Higgs" model~\cite{Chacko:2005pe,Craig:2015pha}, although others have been considered \cite{Barbieri:2005ri,Foot200775,Foot:1999ph,DELAGUILA1985237,Foot:1996hp}.  The Twin Higgs model contains a full or partial copy of the SM gauge group and particle content, with couplings in the two sectors related by a discrete ${\mathbb Z}_2$ symmetry. The particle content in each sector, usually denoted $A$ and $B$, transforms under its own gauge group and is sterile with respect to the other sector. An attractive feature of the Twin Higgs model is that it provides a solution to the little hierarchy problem without requiring new particles charged under the SM gauge group, at least below the cutoff of the effective theory. In this model, the Higgs field is a pseudo-Nambu Goldstone boson (PNGB) associated with spontaneous breaking of an approximate global $SU(4)$ symmetry. A twin ${\mathbb Z}_2$ symmetry is introduced to constrain the form of corrections to the PNGB Higgs potential, allowing natural electroweak symmetry breaking with no quadratically divergent corrections to the Higgs mass at one-loop level. 

In this paper, we construct a 3+3 neutrino model within the Twin Higgs framework, although many of our phenomenological studies can be applied to other models with similar flavor structures. Two higher dimensional operators turn out to be relevant for the neutrino sectors. The first operator is dimension-five and respects both the ${\mathbb Z}_2$ and $SU(4)$ symmetries. After the Higgs fields develop their  vacuum expectation values (VEV's), three out of the total six neutrinos become massive and can be identified as the three sterile neutrinos. Because of ${\mathbb Z}_2$-enforced Yukawa alignment between the two sectors, only one universal mixing angle in addition to the usual Pontecorvo-Maki-Nakagawa-Sakata (PMNS) matrix is required to describe how the three sterile neutrinos interact with the SM charged leptons. This mixing angle $\theta$ is predicted to be the ratio of two Higgs VEV's, $v/f\sim {\cal O}(0.1)$, and has the right order of magnitude to fit the SBL excesses. The second relevant operator is dimension-six, which is ${\mathbb Z}_2$-conserving and $SU(4)$-breaking. It is responsible for coupling the Majoron to neutrinos and additionally for providing mass to the light active neutrinos. We will show that after satisfying various constraints, the three heavy sterile neutrinos can decay into active neutrinos plus one Majoron with the decay distance within the experimental lengths. In what follows, we will analyze oscillation and decay of Dirac and Majorana sterile neutrinos within the context of this 3+3 ``Twin Neutrino" model. We will show that promptly decaying sterile neutrinos in this model can provide a good fit to the LSND and MiniBooNE anomalies.

%%%%%%%%%%%%%%%%%%%%%%%%%%%%%%%%
\section{The Twin Neutrino Model}
\label{sec:model}
%%%%%%%%%%%%%%%%%%%%%%%%%%%%%%%
Motivated by the Twin Higgs model, we consider a global non-Abelian $SU(4)$ symmetry in the electroweak parts of both the SM and twin sectors. The two Higgs doublets $H_A$ and $H_B$, which transform under $SU(2)_A\times U(1)_A$ and $SU(2)_B\times U(1)_B$ gauge symmetries, can be grouped together as a quadruplet of $SU(4)$: ${\cal H} = (H_A, H_B)^T$. At the minimum of its $SU(4)$ invariant potential, the quadruplet develops a VEV of $\langle {\cal H} \rangle = (0, 0, 0, f)^T$, spontaneously breaking $SU(4)$ down to its $SU(3)$ subgroup. As a result, there are seven Nambu-Goldstone-bosons (NGB's) in the low energy theory below the cutoff $\Lambda \sim 4\pi f$. Turning on electroweak gauge interactions in both sectors, the quadruplet VEV breaks the twin electroweak gauge symmetry $SU(2)_B \times U(1)_B$ to a single $U(1)$ with three NGB's eaten by the three massive gauge bosons $W^\pm_B$ and $Z_B$.~\footnote{An additional Higgs mechanism may be required to provide the twin photon mass.} The remaining four NGB's can be identified as the SM Higgs doublet, $H$, which acquire mass and become PNGB's after turning on $SU(4)$ breaking gauge or Yukawa interactions.

The little hierarchy problem can be alleviated by imposing an additional $\mathbb{Z}_2$ symmetry between the two sectors which forces all couplings to be the same. This is because in the gauge sector, the one loop corrections to
the Higgs mass which are quadratic in $\Lambda$ have the form $9\Lambda^2/(64\pi^2)(g_A^2  H_A H_A^\dagger + g_B^2 H_B H_B^\dagger) = 9\Lambda^2/(64\pi^2) g^2 {\cal H}{\cal H}^\dagger$ and are independent of the PNGB Higgs field. In addition, the logarithmically divergent part contributes to the coefficient of the $SU(4)$-breaking operator $\kappa\,(|H_A|^4 + |H_B|^4)$ at the order of $g^4/(16\pi^2) \log{(\Lambda/g f)}$, so the Higgs field mass is generically suppressed compared to the VEV $f \sim 1$~TeV. To obtain the lighter Higgs boson mass at 125 GeV, the coefficient is needed to be around 1/4, which suggests additional $SU(4)$ breaking terms in the scalar potential. Minimizing the potential for the two Higgs doublets with small ${\mathbb Z}_2$-breaking terms, the ratio of the two VEV's is
\beqa
\frac{\langle H_A \rangle}{\langle H_B \rangle} = \frac{v}{f} \sim {\cal O}(0.1) \,,
\eeqa
with the electroweak VEV, $v = 246$~GeV. Later we will show that this ratio will be crucial to determine the fermion mass spectrum in the twin sector.

The fermion Yukawa couplings explicitly break the global $SU(4$) symmetry. The ${\mathbb Z}_2$ twin parity is required to ensure that there are no corrections to the SM Higgs mass proportional to $\Lambda^2$ at the one-loop level. Therefore, we keep the Yukawa couplings in both sectors to be exactly the same, both for charged leptons and neutrinos. For the charged leptons, the ${\mathbb Z}_2$-invariant and $SU(4)$-breaking Yukawa couplings in the Lagrangian are
\beqa 
y_e^{ij} \left( H_A \overline{L}_{A\, i} E_{A\, j} +  H_B \overline{L}_{B\, i} E_{B\, j} \right)  \, + \, \mbox{h.c.} \,. 
\eeqa
Here, the indexes ``$i, j = 1,2, 3$" denote lepton flavors. After inputting the scalar VEV's, we have the three twin charged-lepton masses exactly proportional to the SM ones: $m_{e^i_A}/m_{e^i_B} = v/f$. For instance, the twin electron is anticipated to have a mass of $m_{e_B} = {\cal O}(5~\mbox{MeV})$. 

%%%%%%%%%%%%%%%%%%%%%%%%%%%%%%%%
\subsection{Majorana Neutrinos}
\label{sec:model-Majorana}
%%%%%%%%%%%%%%%%%%%%%%%%%%%%%%%
In the neutrino sector, we will consider both Majorana and Dirac neutrino cases and will only focus on the spectrum with normal ordering, which is preferred for the Majorana case. In this subsection, we first study the Majorana neutrinos for both SM and twin sectors. Different from the charged-lepton sector, we can have the following ${\mathbb Z}_2$-invariant and $SU(4)$-conserving Majorana mass operators
\beqa
\frac{y_\nu^{ij}}{\Lambda_{S}} \, {\cal L}^T_{i}  \widetilde{{\cal H}} \,{\cal C} \, \widetilde{{\cal H}}^T  {\cal L}_{j}  + \mbox{h.c.} \,
\label{eq:majorana-mass-operator-1}
\eeqa
Here, ${\cal L} \equiv (L_A, L_B)$ and $\widetilde{H}^T \equiv (\tilde{H}^T_A, \widetilde{H}^T_B)$ with $\widetilde{H}^T_{A, B} \equiv -i \sigma_2 H_{A,B}^T$. The cutoff, $\Lambda_S$, could be related to the some heavy right-handed neutrino masses to realize the See-Saw mechanism. After Higgs doublets get their VEV's, the linear combinations $(v\,\nu_{A, L}^i + f\,\nu_{B, L}^i)$ are massive and are approximately the three sterile neutrino states. To provide mass for other combinations, we introduce the following ${\mathbb Z}_2$-invariant and $SU(4)$-breaking dimension-six operator
\beqa
\frac{y_\nu^{ij}}{\Lambda_{\phi}^2 } \,\phi\, L^T_{A\,i}  \widetilde{H}_A \,{\cal C} \, \widetilde{H}^T_B  {L}_{B\,j}   
 + \mbox{h.c.} \,
\label{eq:majorana-mass-operator-2}
\eeqa
 Here, the new gauge-singlet scalar $\phi$ carries both SM and twin lepton numbers. Furthermore, one could define a discrete symmetry in the twin sector with $\phi \leftrightarrow - \phi$, $L_A \leftrightarrow L_A$ and $L_B \leftrightarrow -L_B$, such that additional operators for $\phi$ coupling to only SM leptons are forbidden. This discrete symmetry is important for our later discussion of $\phi$-related phenomenology. The Yukawa couplings are chosen to be identical for Eq.~(\ref{eq:majorana-mass-operator-1}) and Eq.~(\ref{eq:majorana-mass-operator-2}), which could originate from a UV theory at a much higher than TeV scale.
 
\begin{figure}[th!]
\begin{center}
\includegraphics[width=0.45\textwidth]{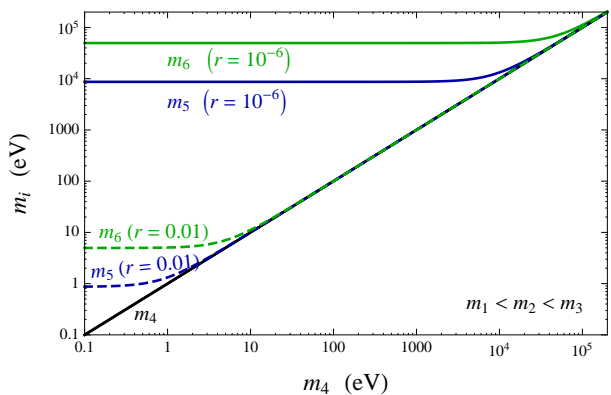}
\caption{The twin neutrino masses as a function of the lightest twin neutrino mass. The normal ordering mass spectrum is assumed for the active neutrinos.}
\label{fig:mass-spectrum}
\end{center}
\end{figure}

 After $\phi$ develops a VEV with $\langle \phi \rangle = f_\phi$,~\footnote{We take the VEV of $\phi$ to be a real number. Its complex phase is physical, but will not change our phenomenological study later.} the $6\times 6$ neutrino mass matrix is 
 \beqa
 M &=& \left[
\begin{array}{cc}
\frac{v^2}{\Lambda_S}  &   \frac{v\,f}{\Lambda_S}(1 + \frac{f_\phi \Lambda_S}{\Lambda_\phi^2})  \vspace{1mm} \\
\frac{v\,f}{\Lambda_S}(1 + \frac{f_\phi \Lambda_S}{\Lambda_\phi^2})  &  \frac{f^2}{\Lambda_S}  
\end{array}
 \right] \otimes y_\nu^{3\times3} \, \\
 &=& {\cal U}^T \mbox{diag}\left(m_{\nu_a^1}, m_{\nu_a^2}, m_{\nu_a^3}, m_{\nu_s^1}, m_{\nu_s^2}, m_{\nu_s^3} \right) {\cal U}  \,.
 \eeqa
In the leading order of  $f_\phi \Lambda_S/\Lambda_\phi^2 \ll 1$ and $v/f \ll 1$, the three heavy (sterile) neutrino masses are
 \beqa
 m_{\nu_s^i} \approx \bar{y}^i_\nu \frac{f^2}{\Lambda_S} \,,
 \eeqa
 where $\bar{y}^i_\nu$ is the eigenvalue of the Yukawa matrix $y_\nu$. Because of flavor alignment in the SM and twin sectors, the ratios of the neutrino masses satisfy
 \beqa
 r \equiv \frac{m_{i}}{m_{i+3}} = \frac{m_{\nu_a^i}} {m_{\nu_s^i}} \approx 2\,\frac{f_\phi \Lambda_S}{\Lambda_\phi^2} \frac{v^2}{f^2} \,,
 \eeqa
to the leading order of the small parameter, $f_\phi \Lambda_S/\Lambda_\phi^2 \ll 1$, in our model. For a normal ordering mass spectrum of active neutrinos, $m_1 < m_2 < m_3$, and from Ref.~\cite{Agashe:2014kda}, we have $\Delta m^2_{21} \approx 7.54\times 10^{-5}$~eV$^2$ and $\Delta m^2 \equiv \Delta m^2_{31} - \Delta m^2_{21}/2 \approx 2.43\times 10^{-3}$~eV$^2$. The twin neutrino masses are shown in Fig.~\ref{fig:mass-spectrum} for two different values of $r$. 

In the charged-lepton mass-eigenstate basis, the diagonalization unitary matrix ${\cal U}$ can be written in terms of a tensor product
\beqa
{\cal U}^{6\times 6} = O^{2\times 2} \otimes U^{3\times3}_{\rm PMNS} \,,
\eeqa
where $U^{3\times3}_{\rm PMNS}$ is the PMNS matrix in the SM (the experimental values are taken from Ref.~\cite{Agashe:2014kda}) and $O^{2\times 2}$ is a rotation matrix  
\beqa
O^{2\times 2} =  \left(
\begin{array}{cc}
\cos{\theta}  &   \sin{\theta}  \\
 - \sin{\theta}  &  \cos{\theta}  
\end{array} 
 \right)  \,. 
\eeqa
To the leading order of $f_\phi \Lambda_S/\Lambda_\phi^2\ll 1$, the new mixing angle between the SM and twin neutrino sectors is
\beqa
\theta \approx \frac{ v} { f } = {\cal O}(0.1) \,.
\label{eq:O-matrix}
\eeqa
Consequently, the three active neutrinos ($\nu_a^i$) interact with the SM charged leptons with a strength proportional to $\cos{\theta}$, while the three sterile neutrinos ($\nu_s^i$) have interactions suppressed by $\sin{\theta}$. {\it The sterile neutrino interaction strengths with SM charged leptons are therefore related to the fine-tuning problem for the SM Higgs boson.}

In this model, there is a PNGB or Majoron~\cite{Chikashige:1980qk,Chikashige:1980ui,Gelmini:1980re} associated with the global symmetry breaking of $U(1)_{L_A}\times U(1)_{L_B} \rightarrow U(1)_L$. The relevant Yukawa couplings for the Majoron particle, $\varphi$, defined as $\phi \equiv f_\phi e^{i \varphi/f_\phi}$, in our models is flavor diagonal and are
\beqa
\lambda_{a s}^i \,i\varphi\, \nu_{a, L}^{i\,T} {\cal C} \nu_{s,L}^i &\equiv& \bar{y}_\nu^i \, \frac{v f}{2\Lambda_\phi^2} \,i\varphi\,\nu_{a,L}^{i\,T} {\cal C} \nu_{s,L}^i  \,, \nonumber \\
\lambda_{a a}^i \,i\varphi\, \nu_{a,L}^{i\,T} {\cal C} \nu_{a,L}^i &\equiv& \theta\,\bar{y}_\nu^i \, \frac{v f}{2\Lambda_\phi^2} \,i\varphi\,\nu_{a,L}^{i\,T} {\cal C} \nu_{a,L}^i  \,.
\label{eq:transition-Yukawa}
\eeqa
The sterile neutrinos are not stable particles and have the decay widths of 
\beqa
\Gamma[\nu_s^i]&=&2\,\Gamma[\nu_s^i \rightarrow \nu_a^i (\bar{\nu}_a^i ) + \varphi] 
\nonumber \\
&\approx& \frac{(\lambda_{as}^i)^2\,m_{\nu_s^i}}{4\pi}  
= \frac{(\lambda_{as}^3)^2}{4\pi} \frac{m_{\nu_s^i}^3}{m_{\nu_s^3}^2} \,.
\eeqa
In our numerical study later, we will focus on the parameter region of $\lambda_{as}^3 = {\cal O}(10^{-2})$ or $\bar{y}^3_\nu = {\cal O}(10^{-2})$ and $\Lambda_\phi = {\cal O}(10\,\sqrt{v\,f})={\cal O}(\mbox{TeV})$. 

Since the mass operators in Eq.~(\ref{eq:majorana-mass-operator-1}) explicitly break the $U(1)_L$ symmetry, we anticipate a non-zero mass for the Majoron filed, $\varphi$. The one-loop diagram mediated by $L_A$ and $L_B$ generates a mass for $\varphi$ 
\beqa
m^2_\varphi &\sim& \frac{1}{16\pi^2} \frac{v^4 f^4}{\Lambda_\phi^4 \Lambda_S^2} \mbox{Tr}(y_\nu y_\nu^\dagger y_\nu y_\nu^\dagger)  \nonumber \\
&\sim& \frac{1}{16\pi^2} \, (\lambda^3_{as} )^2 \, \theta^2\,m_{\nu_s^3}^2 \, 
 \,.
\eeqa
For a normal ordering neutrino mass spectrum, we have $m_\varphi \approx 1$~eV for $m_{\nu_s^3} \approx 10$~keV, $\theta =0.1$ and $\lambda^3_{as} \sim 0.01$. The Majoron decay width is 
\beqa
\sum_i \Gamma(\varphi \rightarrow 2 \nu_a^i) = \sum_i \frac{\theta^2 \,  (\lambda^3_{as} )^2}{4\pi} \, \frac{m_{\nu_a^i}^2}{m_{\nu_a^3}^2} m_\varphi \, \Theta(m_\varphi - 2 m_{\nu_a^i}) \,.
\eeqa
For a normal ordering neutrino mass spectrum, $m_\varphi \approx 1$~eV, $\lambda^3_{as} \sim 0.01$ and $\theta=0.1$, the decay width at rest is $\Gamma_\varphi \approx 8\times 10^{-8}$~eV.

%%%%%%%%%%%%%%%%%%%%%%%%%%%%%%%%
\subsection{Dirac Neutrinos}
\label{sec:model-Dirac}
%%%%%%%%%%%%%%%%%%%%%%%%%%%%%%%
For the Dirac neutrino case, one need to introduce an additional set of right-handed neutrinos in both the SM and twin sectors. The $\mathbb{Z}_2$-invariant and $SU(4)$-conserving Dirac mass operators are
\beqa
y_\nu^{ij} \, \widetilde{{\cal H}}\, \overline{{\cal L}} (\nu_{A, R} + \nu_{B, R} ) \,+\,\mbox{h.c.}
\eeqa
Furthermore, the following $\mathbb{Z}_2$ and $SU(4)$-breaking dimension-five operator is introduced to provide light neutrino masses and decay couplings for the heavy neutrinos
\beqa
\frac{y_\nu^{ij}}{\Lambda_\phi} \phi \, \widetilde{H}_A \,\overline{L}_A\, \nu_{B, R}
  \,+\,\mbox{h.c.}
\eeqa

 After $\phi$ develops a VEV with $\langle \phi \rangle = f_\phi$, the $6\times 6$ neutrino mass matrix is 
 \beqa
 M &=& \frac{1}{\sqrt{2}}\, \left[
\begin{array}{cc}
v  &  v (1 + \frac{f_\phi}{\Lambda_\phi}) \vspace{1mm} \\
f  &  f
\end{array}
 \right] \otimes y_\nu^{3\times3}  \, \\
 &=& {\cal U}^T \mbox{diag}\left(m_{\nu_a^1}, m_{\nu_a^2}, m_{\nu_a^3}, m_{\nu_s^1}, m_{\nu_s^2}, m_{\nu_s^3} \right) {\cal W}  \,.
 \eeqa
 with the left-handed rotation matrix as
 \beqa
{\cal U}^{6\times 6} = O^{2\times 2}_L \otimes U^{3\times3}_{\rm PMNS} \,.
 \eeqa
Using the same parametrization in Eq.~(\ref{eq:O-matrix}) and in the limit of $f_\phi \ll \Lambda_\phi$, we still have $\theta \approx v/f = {\cal O}(0.1)$. The mass ratios for this Dirac neutrino model is
 \beqa
 r \equiv \frac{m_{i}}{m_{i+3}} = \frac{m_{\nu_a^i}} {m_{\nu_s^i}} \approx \frac{1}{\sqrt{2} }\,\frac{f_\phi}{\Lambda_\phi} \frac{v}{f} \,. 
 \eeqa

In the Dirac neutrino model, we also have a PNGB associated with the symmetry breaking of $U(1)_{L_A}\times U(1)_{\nu_{B, R}} \rightarrow U(1)_L$. The couplings of the Majoron, $\varphi$, parametrized by $\phi \equiv f_\phi e^{i \varphi/f_\phi}$ are
\beqa
\lambda^i_{as}\, i\varphi \,\overline{\nu^i_{a, L}} \, \nu^i_{s, R} &\equiv& \bar{y}^i_\nu \, \frac{v}{2\,\Lambda_\phi}\,i\varphi \,\overline{\nu^i_{a, L}} \, \nu^i_{s, R} \,, \\
\lambda^i_{aa}\, i\varphi \, \overline{\nu^i_{a, L}} \, \nu^i_{a, R} &\equiv& \bar{y}^i_\nu \, \frac{v}{2\,\Lambda_\phi}\,i\varphi \,\overline{\nu^i_{a, L}}\, \nu^i_{a, R} \,. 
\eeqa
The sterile neutrino decay widths are
\beqa
\Gamma[\nu_s^i \rightarrow \nu_a^i + \varphi] \approx \frac{(\lambda_{as}^i)^2}{32\pi} m_{\nu_s^i} 
= \frac{(\lambda_{as}^3)^2}{32\pi} \frac{m_{\nu_s^i}^3}{m_{\nu_s^3}^2} \,.
\eeqa
In our model, the active neutrinos from sterile neutrino decays are left-handed.

%%%%%%%%%%%%%%%%%%%%%%%%%%%%%%%%
\section{Constraints from unitary, meson decays and neutrinoless double beta decay}
\label{sec:constraints}
%%%%%%%%%%%%%%%%%%%%%%%%%%%%%%%
For both the Majorana and Dirac neutrino models, we have the model parameters: $m_4$, $\Gamma_4$ (related to the coupling $\lambda_{as}^3$), $\theta$ and $r$. In this section, we study the existing constraints on our model parameters from unitarity of the active neutrino mixing matrix, neutrinoless double beta decay and meson decays.  

The mixing between active and twin neutrinos reduces the couplings of active neutrinos to charged leptons in the SM. The $6\times 6$ mixing matrix, $\cal U$, is a unitary matrix in our model, but the $3\times 3$ mixing matrix, ${U}$, is not unitary and has the normalization property of $\sum_{i=1}^3 |U_{\ell i}|^2 = \cos^2{\theta}$.  Using the results in Ref.~\cite{Parke:2015goa} and neglecting the effects of sterile decay products, we have found that 
\beq
\sin\theta \lesssim 0.20  \,,
\eeq
at $2\sigma$ C.L. 

\begin{figure}[th!]
\begin{center}
\includegraphics[width=0.45\textwidth]{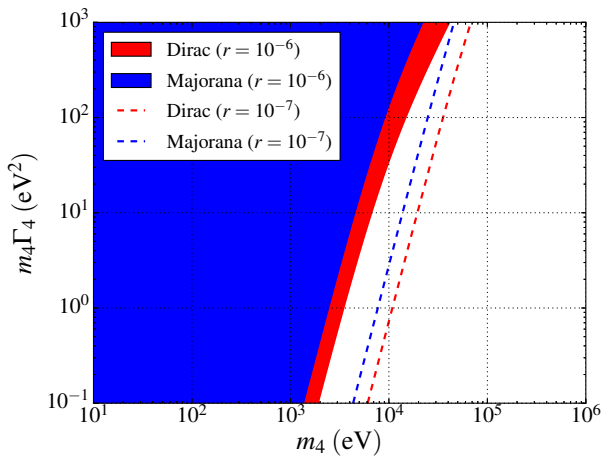}
\caption{The constraints on our model parameters from the $e-\mu$ universality of pion decays. The Dirac phase $\delta_{CP}=0$.}
\label{fig:constraints-meson-decay}
\end{center}
\end{figure}

There are additional bounds on the sterile neutrino decay widths from the pion and kaon three-body decay into the new Majoron state and the electron-muon universality tests of their total leptonic widths (see Ref.~\cite{deGouvea:2015euy} for a recent summary). Using the analysis in Ref.~\cite{Barger:1981vd} and the  measurement of $\pi\rightarrow e\nu$ branching ratio~\cite{Britton:1993cj,Aguilar-Arevalo:2015cdf}, the predicted deviation from $e-\mu$ universality is $R_\pi = 1 + 157.5 (g^2)_{ee}$ with the experimental value of  $R_\pi=0.9931\pm 0.0049$ (the updated value of $R_\pi=0.9993\pm 0.0024$ provides a similar bound), so the bound on our model parameters is
\beqa
(g^2)_{ee} = \,c\,\sum_{i=1}^3 U_{e i} \frac{32\pi\,\Gamma_{\nu_s^i \rightarrow \nu_a^i + \varphi} }{m_{\nu_s^i}} U^T_{i e} <  3.0 \times 10^{-5} \,, 
\eeqa
at 90\% C.L. Here, $c=1(2)$ for the Majorana(Dirac) neutrino case.  In Fig.~\ref{fig:constraints-meson-decay}, we show the constraints on our model parameters in the $m_4 \, \Gamma_4$ and $m_4$ plane. 

\begin{figure}[th!]
\begin{center}
\includegraphics[width=0.45\textwidth]{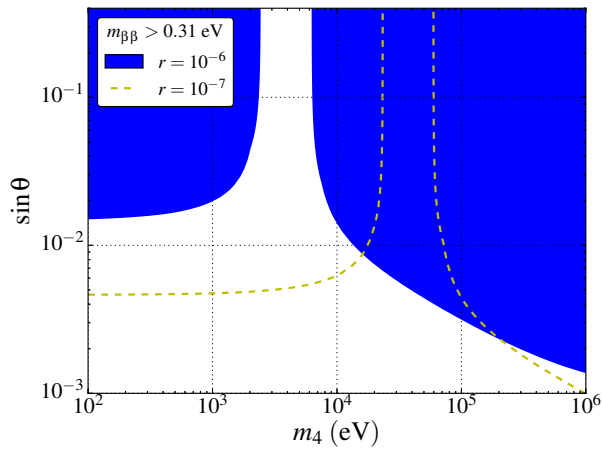}
\caption{Constraints on $m_4$ and $\sin{\theta}$ from $0\nu 2\beta$ decay with $|\langle m_{\beta\beta} \rangle| < 0.31$~eV~\cite{Guzowski:2015saa}. We have chosen a normal ordering mass spectrum for the three active neutrinos and the Dirac phase $\delta_{CP}=0$.}
\label{fig:constraints-neutrinoless}
\end{center}
\end{figure}

For the Majorana neutrino case, searches for the neutrinoless double beta decay $(\beta\beta)_{0\nu}$ can also impose bounds on our model parameter space. The amplitude of $(\beta\beta)_{0\nu}$ is proportional to the effective Majorana mass
\beqa
|\langle m_{\beta\beta} \rangle| \equiv \left| \sum_{i=1}^{6} m_i\,{\cal U}^2_{ei} \right|
= (r \cos^2{\theta} + \sin^2{\theta}) \left| \sum_{i=1}^{3} m_{3+i}\,U^2_{ei} \right| \,.
\eeqa
The $CP$-violating phases can affect the predicted effective Majorana mass. In our model, we have identical Majorana and Dirac phases for the SM and twin sectors. In Fig.~\ref{fig:constraints-neutrinoless}, we show the allowed parameter space by allowing arbitrary Majorana phases but fixed Dirac phase of $\delta_{CP}=0$. 
We did not find additional processes like $0\nu 2\beta \varphi$ or $0\nu 2\beta 2\varphi$ provide more stringent bounds than from $0\nu 2\beta$ and meson decays. 

Because of the fairly large interaction strength ($\lambda_{as}^3 \sim 0.01$) among sterile neutrinos, active neutrinos and the Majoron particle, both sterile neutrinos and the Majoron particles stay within the supernova core. Our model does not have an additional energy loss problem for supernova SN1987A. For sterile neutrinos with a mass below around 1~MeV, they contribute to additional relativistic degrees of freedom and are constrained from BBN~\cite{Barger:2003zg,Ade:2015xua}. However, to have a conclusive statement, additional analysis or non-standard cosmology should be taken into account. We do not explore these directions here.

%%%%%%%%%%%%%%%%%%%%%%%%%%%%%%%%
\section{Neutrino Appearance and Disappearance}
\label{sec:neutrino-formula}
%%%%%%%%%%%%%%%%%%%%%%%%%%%%%%%
In this section, we write down general neutrino appearance and disappearance formulas for our model. Since we have both oscillation and decay, we will keep both effects for neutrino appearance and disappearance. For the short-baseline experiments, the differential probability for a neutrino of flavor $\alpha$ with energy $E_{\nu_\alpha}$ converting into a neutrino of flavor $\beta$ with energy in the interval of $(E_{\nu_\beta}, E_{\nu_\beta} + dE_{\nu_\beta})$ is~\cite{Lindner:2001fx,PalomaresRuiz:2005vf} 
\beqa
\hspace{0cm}\frac{dP_{\nu_\alpha \rightarrow \nu_\beta}(E_{\nu_\alpha})}{dE_{\nu_\beta}}  &=&  \left| \sum_{i=1}^6 {\cal U}_{\beta i} {\cal U}^*_{\alpha i} e^{ - i \frac{m_i^2 L}{2 E_{\nu_\alpha}} - \frac{m_i\Gamma_i L}{2E_{\nu_\alpha}}  }  \right|^2 \,\nonumber  \\
&&\hspace{-3cm} \times\, \delta(E_{\nu_\alpha}-E_{\nu_\beta})
+ W_{E_{\nu_\alpha}\rightarrow E_{\nu_\beta}} \nonumber  \nonumber \\
&& \hspace{-3cm} \times\, \int^L_0 dL^\prime\, \frac{c}{2} \left|  
\sum_{j=4}^6 {\cal U}_{\beta (j-3)} {\cal U}^*_{\alpha j} \sqrt{ \frac{m_j \Gamma_j}{E_{\nu_\alpha} }} e^{ - i \frac{m_j^2 L^\prime}{2 E_{\nu_\alpha}} - \frac{m_j \Gamma_j L^\prime}{2 E_{\nu_\alpha}}  }
\right|^2 \,.
\label{eq:decay-plus-oscillation}
\eeqa
Here, the neutrino decay widths, $\Gamma_i=0$ for $i=1,2,3$ and $\Gamma_i\neq 0$ for $i=4,5,6$, are defined in the rest frame. $c=1(2)$ for the Majorana(Dirac) neutrino case. For the case of neutrino goes to neutrino, the energy spectrum function is $W_{E_{\nu_\alpha}\rightarrow E_{\nu_\beta}} = 2 E_{\nu_\beta}/E_{\nu_\alpha}^2 \,\Theta(E_{\nu_\alpha}-E_{\nu_\beta})$. For the Majorana model, the helicity-flip formula for $\nu_\alpha \rightarrow \overline{\nu}_\beta$ has only the second term (the decaying part) with a different energy spectrum function, $W_{E_{\nu_\alpha}\rightarrow E_{\overline{\nu}_\beta}} = 2 (E_{\nu_\alpha} - E_{\overline{\nu}_\beta})/E_{\nu_\alpha}^2 \,\Theta(E_{\nu_\alpha}-E_{\overline{\nu}_\beta})$. For the case with initial anti-neutrinos, one should replace the elements of ${\cal U}$ by their complex conjugates. 

In our numerical studies, we will focus on the pure oscillation case as well as the case where the decay effect dominates. For the pure oscillation case with $\Gamma_i =0$ and in the limits of $m_{4,5,6} \gg m_{1,2,3}$ and $|\Delta m^2_{12}|, |\Delta m^2_{23}| \ll E/L$ (the short-baseline approximation), the neutrino appearance probability is 
\beqa
P(\nu_\alpha \rightarrow \nu_\beta)&=& 4 \sin^4{\theta} \sum_{j=2}^3 \sum_{k=1}^{j-1} |U_{\beta j}U^*_{\alpha j} U^*_{\beta k} U_{\alpha k} | \nonumber \\ 
&&\hspace{-1.5cm}\times\,\sin{ \left(\frac{\Delta m^2_{jk} L}{2\,r^2 E }\right) } \sin{\left(\phi_{\beta \alpha j k} -  \frac{\Delta m^2_{jk} L}{2\,r^2 E }\right) } \,,
\label{eq:oscillation-app}
\eeqa
for $\alpha\neq \beta$. Here, the phase $\phi_{\beta \alpha j k} \equiv \mbox{arg}(U_{\beta j}U^*_{\alpha j} U^*_{\beta k} U_{\alpha k} )$. The formula  for the anti-neutrino case can be obtained by a replacement of $\phi \rightarrow - \phi$.  In our model, only a single Dirac $CP$ phase enters both sectors. In the small mixing angle limit of $\theta_{13}\ll 1$ and $\delta_{CP}$ order of unity, we have $\phi_{e \mu 2 1} = {\cal O}(\theta_{13} \,\sin{\delta_{CP}} )$, $\phi_{e \mu 31} = \phi_{e \mu 32} = - \delta_{CP} + {\cal O}(\theta_{13})$. So, a large $CP$-violating phase can affect the (anti-)neutrino appearance probabilities. The disappearance probability is independent of $CP$-violating phase and has
\beqa
P(\nu_\alpha \rightarrow \nu_\alpha) = 1 - \sin^2{(2\theta)} \sum_{j=1}^3 |U_{\alpha j}|^2  \sin^2{ \left(\frac{\Delta m^2_{j+3, 1} L}{2\, E }\right) }  \,.
\label{eq:oscillation-disapp}
\eeqa
Comparing Eqs.~(\ref{eq:oscillation-app}) and (\ref{eq:oscillation-disapp}), one can see that the appearance probability is suppressed by $\sin^4{\theta}$, while the disappearance is only suppressed by $\sin^2{\theta}$. We will later show because of this fact it is challenging to only use oscillation to explain LSND and MiniBooNE anomalies. 

For the case in which sterile neutrino decay effects are dominant and in the limit of $|\Delta m^2_{45, 56, 46}| \gg \Gamma_j m_j$, the appearance probability is
\beqa
\frac{dP_{\nu_\alpha \rightarrow \nu_\beta}(E_{\nu_\alpha})}{dE_{\nu_\beta}}&=& \frac{c}{2} \sin^2{\theta}\;W_{E_{\nu_\alpha}\rightarrow E_{\nu_\beta}} \, 
\nonumber \\
&&  \hspace{-3.5cm} \times \, \sum_{j=1}^3 |U_{\beta j}|^2 |U_{\alpha j}|^2 \left( 1 - e^{- \frac{m_{j+3} \Gamma_{j+3} L }{E_{\nu_\alpha} } }   \right) + {\cal O}(\sin^4{\theta}) \,.
\label{eq:decay-appearance}
\eeqa
For the Majorana model, the helicity-flip formula for $\nu_\alpha \rightarrow \overline{\nu}_\beta$ is similar but using $W_{E_{\nu_\alpha}\rightarrow E_{\overline{\nu}_\beta}}$. 
The disappearance has contributions from both terms in Eq.~(\ref{eq:decay-plus-oscillation}) and is
\beqa
\frac{dP_{\nu_\alpha \rightarrow \nu_\alpha}(E_{\nu_\alpha})}{dE_{\nu_\alpha^\prime}}&=& \nonumber \\
&&  \hspace{-3.0cm} \left[ 1 - 2 \sin^2\theta  +\sin^2\theta \sum_{j=1}^3 |U_{\alpha j}|^2 e^{- \frac{m_{j+3} \Gamma_{j+3} L }{E_{\nu_\alpha} } } \right]\delta(E_{\nu_\alpha} - E_{\nu_\alpha^\prime})  \nonumber \\
&&\hspace{-3.0cm} + W_{E_{\nu_\alpha}\rightarrow E_{\nu_\alpha^\prime}} \,\times\,\frac{c}{2} \sin^2{\theta} \, \sum_{j=1}^3 
|U_{\alpha j}|^4 \left( 1 -  e^{- \frac{m_{j+3} \Gamma_{j+3} L }{E_{\nu_\alpha} } }  \right) \,,
\eeqa
ignoring additional terms suppressed by $\sin^4{\theta}$. One can see that for the decay case both appearance and disappearance are suppressed by the same power of $\sin{\theta}$. This fact will make the decay model a better fit to LSND and MiniBooNE data.

%%%%%%%%%%%%%%%%%%%%%%%%%%%%%%%%
\section{Fit to LSND, MiniBooNE and SciBooNE data}
\label{sec:neutrino-data}
%%%%%%%%%%%%%%%%%%%%%%%%%%%%%%%
In this section, we will consider both cases for interpreting short-baseline experimental data with or without sterile neutrino decays. For the first case without sterile neutrino decays, the energy spectrum of the neutrinos near the far detector follows the initial injected neutrino spectrum. On the other hand, for the second case with sterile neutrino decays, additional care should be taken to account for the energy spectrum change. 

%%%%%%%%%%%%%%%%%%%%%%%%%%%%%%%%
\subsection{Oscillation without decay}
\label{sec:neutrino-data-oscillation}
%%%%%%%%%%%%%%%%%%%%%%%%%%%%%%%
To interpret the LSND event excess of anti-neutrino appearance, $\bar{\nu}_\mu \rightarrow \bar{\nu}_e$~\cite{Aguilar:2001ty} (we will ignore the less significant excess in the neutrino appearance observed by LSND~\cite{Athanassopoulos:1997er}), and to derive the preferred model parameter space, we follow Ref.~\cite{Giunti:2010jt} to account the energy spectrum as well as the conversion of the neutrino energy spectrum to measured positron energy spectrum, $E_{e^+}=E_{\bar{\nu}_e} + m_p - m_n$, from the inverse neutron decay process $\bar{\nu}_e + p \rightarrow n + e^+$. The measured positron energy has the range of $(20~\mbox{MeV}, 60~\mbox{MeV})$. For the simplest $3+1$ sterile neutrino model, we have reproduced the LSND contours in Ref.~\cite{Giunti:2010jt}. 

For the MiniBooNE $\bar{\nu}_\mu \rightarrow \bar{\nu}_e$ and $\nu_\mu \rightarrow \nu_e$ appearance analysis based on initial anti-neutrino and neutrino fluxes~\cite{Aguilar-Arevalo:2013pmq}, we use the data released by MiniBooNe collaboration~\cite{MiniBooNE} to derive the preferred contours in our model parameter space. We combine the two data sets in the (anti-)neutrino energy range of $(200~\mbox{MeV}, 3~\mbox{GeV})$. Again, for the simplest $3+1$ sterile neutrino model, we reproduce the contours in the MiniBooNE publication~\cite{Aguilar-Arevalo:2013pmq}. 

For the constraints from (anti-)neutrino disappearance, we use the combined SciBooNE and MiniBooNE analysis~\cite{Cheng:2012yy} for the anti-neutrino disappearance measurement, which provides more stringent constraints than the ones from the neutrino disappearance measurement~\cite{Mahn:2011ea}. We use the publicly available data~\cite{SciBooNE} to constrain our model parameter space. We have also checked additional constraints from appearance searches by KARMEN~\cite{Armbruster:2002mp} but found them to be less stringent, and we will not report them here.

\begin{figure}[th!]
\begin{center}
\includegraphics[width=0.45\textwidth]{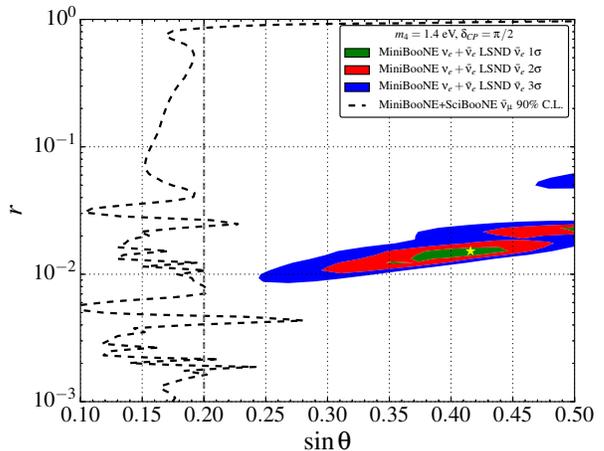}
\caption{The MiniBooNE plus LSND preferred contours for the purely oscillation case in our model. The vertical dashed line at $\sin{\theta} =0.20$ is the constraint line from unitarity of the three active neutrinos. The yellow star is the best fit point of our model.}
\label{fig:oscillation}
\end{center}
\end{figure}

In Fig.~\ref{fig:oscillation}, we show the LSND and MniBooNE preferred contours in terms of our model parameters: $\sin{\theta}$ and $r$ by fixing $m_4 = 1.4$~eV and assuming $\delta_{CP}=\pi/2$, which provides a better fit than no $CP$ violation with $\delta_{CP}=0$. From a three-dimensional parameter scan, we have found the point with the smallest $\chi^2$ at $\sin{\theta}=0.4$, $r=0.011$ and $m_4=1.4$~eV. This is an improvement by $\Delta \chi^2 = 25$ compared to the no oscillation fit. Unfortunately, the disappearance constraints from SciBooNE and MiniBooNE exclude all the $3\sigma$ appearance-data-preferred region. Furthermore, the constraints from unitarity of the three active neutrinos also exclude the LSND and MniBooNE preferred contours.
This can be understood by the formulas in Section~\ref{sec:neutrino-formula}, which show that the appearance probabilities for the pure oscillation case have an additional $\sin^2{\theta}$ with respect to the disappearance probabilities. The tension for appearance and disappearance data is a general feature even for a general $3+2$ global fit~\cite{Kopp:2013vaa,Giunti:2013aea}. 
%%%%%%%%%%%%%%%%%%%%%%%%%%%%%%%%
\subsection{Oscillation with decay}
\label{sec:neutrino-data-decay}
%%%%%%%%%%%%%%%%%%%%%%%%%%%%%%%
For the second case with the decay effects dominant, we need to know more information about the detectors. The first important piece of information is to know whether the sterile neutrinos generated at the source location have already decayed or not.
It can be seen from Eq.~(\ref{eq:decay-appearance}) that in order to have a larger appearance probability, it is preferable to have all sterile neutrinos decay within the experimental lengths. Taking the Lorentz boost into account, this means $\Gamma_4 m_4 > E_\nu/L$. LSND has the distance range of $(26, 34)$~m  and the energy range of $(20, 60)$~MeV; MiniBooNE has the distance around 540~m and the energy range of $(200, 3000)$~MeV; SciBooNE has the distance around 100 m and the energy range of $(300, 1900)$~MeV for the disappearance analysis~\cite{Cheng:2012yy}. Altogether, if $\Gamma_4 m_4 \gtrsim 1$~eV$^2$, the majority of sterile neutrinos have already decayed before reaching the detector. We also note that for large values of $\Gamma_4 m_4$, all sterile neutrinos decay promptly and only $\theta$ and $\Gamma_4 m_4$ are relevant parameters for both appearance and disappearance experiments.

Both LSND and MiniBooNE are able to generate $\nu_\mu$ or $\overline{\nu}_\mu$ initial fluxes. On the other hand, for the appeared electron neutrino or anti-neutrinos in the detectors, the LSND detectors are different from MiniBooNE and SciBooNE. The LSND experiment can distinguish $\overline{\nu}_e$ and $\nu_e$ because after $\overline{\nu}_e$ interacts via inverse beta decay in the mineral oil target of LSND, both a prompt positron and a correlated 2.2~MeV photon from neutron capture appear. This twofold signature is not true for $\nu_e$. On the other hand, the MiniBooNE and SciBooNE can not distinguish $\overline{\nu}$ and $\nu$. In order to compare to the MiniBooNE and SciBooNE data, we need to add both $\overline{\nu}$ and $\nu$. Furthermore, we also note that the quasi-elastic cross sections for $\overline{\nu}$ and $\nu$ interacting with CH$_2$ in MiniBooNE are different. Using the cross sections in GENIE~\cite{Andreopoulos:2015wxa} and for the relevant energy range of $(200, 3000)$~MeV at MiniBooNE, we have 
\beqa
\hspace{-0.2cm} \sigma^{\rm quasi-elastic}_{\nu + \mbox{CH}_2 } > \sigma^{\rm quasi-elastic}_{\overline{\nu} + \mbox{CH}_2} \,,
\label{eq:quasi-elastic}
\eeqa
which is mainly due to $\sigma(\nu n \rightarrow \ell^- p) > \sigma(\overline{\nu} p \rightarrow \ell^+ n)$ from a negative axial-vector form factor~\cite{Formaggio:2013kya}. We will show later that because of different scattering cross sections the Majorana and Dirac models have different features for fitting appearance data.

Before we present our results, we also want to comment on the fact that the LSND or the MiniBooNE measured neutrino transition probability is not simply Eq.~(\ref{eq:decay-plus-oscillation}) for the sterile neutrino decay case. They have the measured probability for each energy bin to be $P^{\rm meas.}_i = (\mbox{data}_i - \mbox{bkgnd}_i)/\mbox{(fully oscillated)}_i$~\cite{Aguilar-Arevalo:2014xrr}. For the full oscillation model, their measured probability matches to the theoretical $L/E_\nu$ oscillation probability. For the decay case at hand, the energy spectra of the initial neutrino flux at the source and the final flux at the detector are different. To compare to the LSND and MiniBooNE $L/E_\nu$ data plots, we use the following probability
\beqa
\hspace{-0.5cm} P^{\rm exp.}_{\rm model}(\nu_\alpha \rightarrow \nu_{\beta} + \overline{\nu}_\beta)(E_{\nu_\beta}) &=& \nonumber \\ 
&& \hspace{-4.7cm}  \frac{\Phi_{\nu_\alpha} \otimes c_1 \frac{dP_{\nu_\alpha \rightarrow \nu_\beta }}{dE_{\nu_\beta} } \otimes  \sigma_{\nu_\beta}  +  \Phi_{\nu_\alpha} \otimes c_2 \frac{dP_{\nu_\alpha \rightarrow \overline{\nu}_\beta }}{dE_{\nu_\beta} } \otimes  \sigma_{\overline{\nu}_\beta} }
{\Phi_{\nu_\alpha} \otimes  \sigma_{\nu_\beta} }  \,.
\label{eq:exp-probability}
\eeqa
Here, the symbol ``$\otimes$" means function convolution. For the Majorana model, we have $c_1=1, c_2=0$ for LSND and $c_1=1, c_2=1$ for MiniBooNE. For the Dirac model, we have $c_1=1, c_2=0$ for both LSND and MiniBooNE. For the initial anti-neutrino experiments, one just makes the interchange of $\nu_{\alpha, \beta} \leftrightarrow \overline{\nu}_{\alpha, \beta}$. In deriving the above equation, we have also made the quasi-elastic approximation with the outgoing electron(positron) energy equal to the incoming neutrino(anti-neutrino) energy. 

Based on Eqs.~(\ref{eq:quasi-elastic})(\ref{eq:exp-probability}), there is an interesting observation about our Majorana model prediction for MiniBooNE. Under the approximation of the same energy spectra of the initial fluxes, $\Phi_{\nu_\alpha} \propto \Phi_{\overline{\nu}_\alpha}$, one has
\beqa
P^{\mbox{\tiny MiniBooNE} }_{\mbox{\tiny Majorana}}(\nu_\alpha \rightarrow \nu_{\beta} + \overline{\nu}_\beta) < P^{\mbox{\tiny MiniBooNE} }_{\mbox{\tiny Majorana}}(\overline{\nu}_\alpha \rightarrow \nu_{\beta} + \overline{\nu}_\beta) \,,
\eeqa
simply from the fact that the neutrino quasi-elastic cross section is larger than the anti-neutrino one. 

\begin{figure}[th!]
\begin{center}
\includegraphics[width=0.45\textwidth]{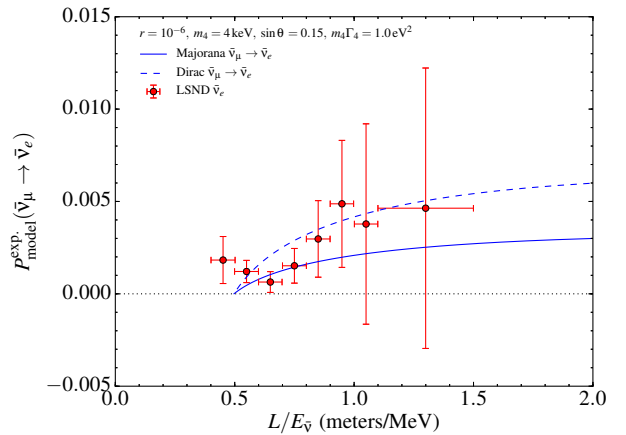}
\caption{A comparison of the decay model prediction and the LSND anti-neutrino appearance data.}
\label{fig:decay-LSND}
\end{center}
\end{figure}
\begin{figure}[th!]
\begin{center}
\includegraphics[width=0.48\textwidth]{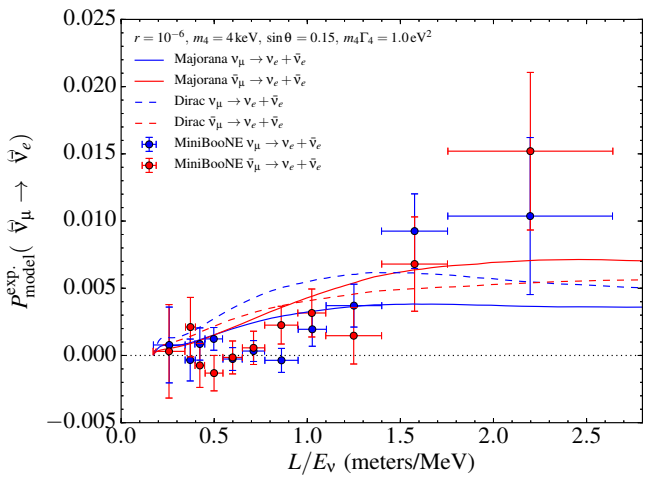}
\caption{A comparison of the decay model prediction and the MiniBooNE neutrino and anti-neutrino appearance data.}
\label{fig:decay-MB}
\end{center}
\end{figure}

In Fig.~\ref{fig:decay-LSND}, we show a comparison of a benchmark model point with $r=10^{-6}$, $m_4 = 4$~keV, $\sin{\theta}=0.15$ and $m_4 \,\Gamma_4 = 1.0$~eV$^2$, to the LSND data. The Dirac model has probabilities higher than the Majorana model simply by a factor of two. Note that the starting points of the model curves are higher than the actual data starting point. This is due to the approximation of using the shortest distance $L=26$~m and the maximal neutrino energy of $52.6$~MeV from muon decays at rest~\cite{Giunti:2010jt}. Similarly for MiniBooNE, in Fig.~\ref{fig:decay-MB} we show both Majorana and Dirac model predictions for the same benchmark model point. As we argued before, the Majorana model has a larger excess for the initial $\overline{\nu}_\mu$ run than the initial $\nu_\mu$ run at MiniBooNE. For the Dirac model, there is also some difference between the initial $\overline{\nu}_\mu$ and $\nu_\mu$ runs, which comes from the slightly different energy spectra for the initial fluxes.  

\begin{figure}[th!]
\begin{center}
\includegraphics[width=0.48\textwidth]{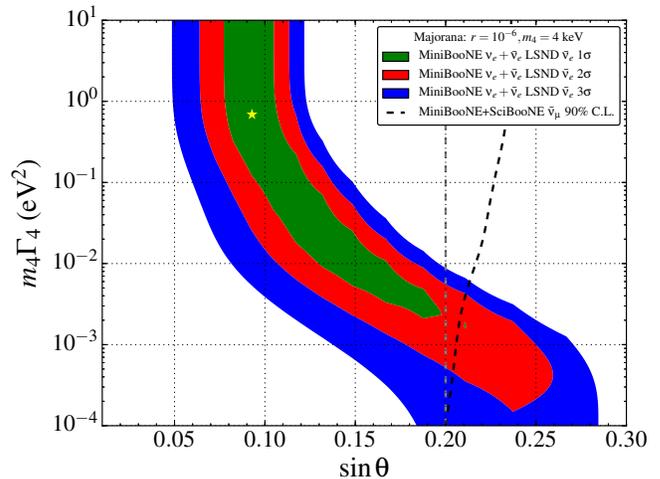}
\caption{The allowed Majorana model parameter space in $\sin{\theta}$ and $m_4 \Gamma_4$ for fixed values of $r$ and $m_4$. The vertical dashed line at $\sin{\theta} =0.20$ is the constraint line (its left side is allowed) from unitarity of the three active neutrinos. The yellow star is the best fit point of our model, which has $\Delta \chi^2 = 22.5$ compared to the no twin neutrino assumption.}
\label{fig:decay-contour}
\end{center}
\end{figure}

In Fig.~\ref{fig:decay-contour} and for the Majorana model, we show the contour plot for the two most relevant model parameters,  $\sin{\theta}$ and $m_4 \Gamma_4$, to fit the MiniBooNE and LSND appearance data. Also shown in this plot is the 90\% C.L. constraints from MiniBooNE plus SciBooNE. Different from the pure oscillation case in Fig.~\ref{fig:oscillation}, the constraint from the disappearance data is not stringent enough to rule out the best fit region for appearance data. For $r=10^{-6}$ and $m_4 = 4$~keV (allowed from the neutrinoless double beta decay constraints in Fig.~\ref{fig:constraints-neutrinoless}), the smallest $\chi^2$ to fit both MiniBooNE (using their Monte Carlo samples) and LSND (using their energy resolution) has $\sin{\theta}=0.097$ and $m_4 \Gamma_4 = 0.55$~eV$^2$. Compared to the fit without new physics, the difference is $\Delta \chi^2 = 22.5$. For two degrees of freedom this means that the background-only fit has a $\chi^2$-probability of $1.3\times 10^{-5}$ (or $4.4 \sigma$) relative to our twin neutrino decay model.  

For the best fit point of the Majorana neutrino model, the six neutrino masses are 0.004~eV, 0.0096~eV, 0.050~eV, 4~keV, 9.56~keV, 50.0~keV. The three sterile neutrino widths are $\Gamma_4=0.0001$~eV, $\Gamma_5=0.0019$~eV and $\Gamma_6=0.27$~eV. The decay Yukawa coupling defined in Eq.~(\ref{eq:transition-Yukawa}) is $\lambda^3_{as} \approx 0.008$. As a result, the Majoron receives a loop-generated contribution to its mass with a value of around 3 eV, which means that the Majoron can decay into all three active neutrinos. For the typical neutrino energy of 30~MeV at LSND (500 MeV at MiniBooNE), the Majoron travels far enough that it can be treated as an invisible particle for both experiments.

\begin{figure}[th!]
\begin{center}
\includegraphics[width=0.48\textwidth]{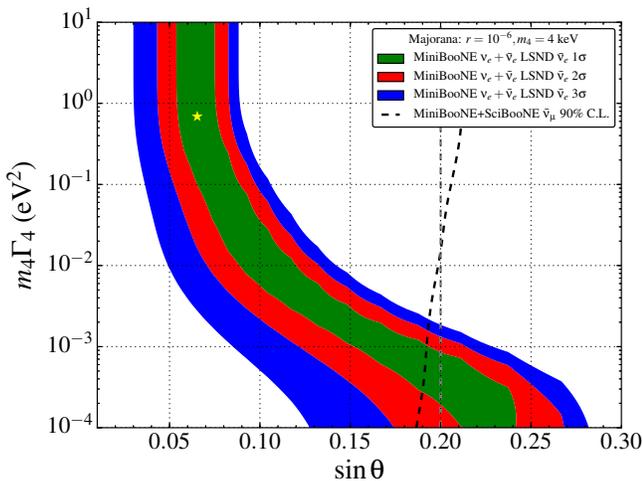}
\caption{The same as Fig.~\ref{fig:decay-contour} but for the Dirac neutrino model. The best fit point has $\Delta \chi^2 = 18.7$ compared to the no twin neutrino assumption.}
\label{fig:decay-contour-dirac}
\end{center}
\end{figure}

For the Dirac neutrino case, we show the allowed parameter region in Fig.~\ref{fig:decay-contour-dirac} for fixed values of $r=10^{-6}$ and $m_4=4$~keV. The best fit point is at $\sin{\theta}=0.073$ and $m_4 \Gamma_4=0.34$~eV$^2$ with $\Delta \chi^2 = 18.7$ (a probability of $8.7\times 10^{-5}$ or $3.9\sigma$) compared to the fit without the three twin neutrinos. The Dirac model provides a slightly worse fit than the Majorana model, because the Majorana model has a higher model prediction for the anti-neutrino appearance probability than the neutrino one. We also note that the anti-neutrino disappearance constraint from MiniBooNE plus SciBooNE for the Dirac model is more stringent than for the Majorana model. This is simply due to the larger scattering cross section of neutrinos than anti-neutrinos. 

%

%%%%%%%%%%%%%%%%%%%%%%%%%%%%%%%%
\section{Discussion and Conclusions}
\label{sec:conclusion}
%%%%%%%%%%%%%%%%%%%%%%%%%%%%%%%
The twin neutrino scenario is very predictive. The neutrino mixing matrix is
fixed by the usual PMNS matrix and the ratio of the Higgs VEV's in the two sectors. Additionally, the signals observed by LSND and MiniBooNE dictates either $m_4^2$ or $m_4\Gamma_4$ to be around $1~{\rm eV}^2$. In the Majorana  sterile neutrino decay scenario, the model also requires a normal ordering mass spectrum for the three active neutrinos to avoid the $0\nu\beta\beta$ decay constraint. Because the SM and twin sectors are closely related by the twin $\mathbb{Z}_2$ symmetry, the number of free parameters is greatly reduced. The next generation neutrino experiments probing the twin sector neutrinos can also provide information about the SM active neutrino sector. Combining the experimental information from both sectors, it is very likely that we can completely determine all the parameters in the neutrino sector.

The decay sterile neutrino scenario provides a novel explanation to the LSND and MiniBooNE anomalies. Depending on whether neutrinos are Dirac or Majorana, one could have either only $\nu_e$ or both $\nu_e$ and $\overline{\nu}_e$ to be decay products of the twin neutrino components in $\nu_\mu$. For the Majorana case, the MiniBooNE $\overline{\nu}_\mu$ run can have a larger excess than the $\nu_\mu$ run. This is because $\nu_e$ as a product of the $\overline{\nu}_\mu$ twin partner has a larger scattering cross section than $\overline{\nu}_e$. Our explanation for the larger $\overline{\nu}_\mu$ run excess does not require $CP$ violation, which is necessary for the pure oscillation explanation.

Future experiments like MicroBooNE~\cite{Soderberg:2009rz} (see also Ref.~\cite{Dutta:2015nlo} for other interesting proposals) may provide decisive tests for the LSND and MiniBooNE excesses. Their results will not just cover our twin neutrino decay scenario, but also the pure oscillation interpretation. We also note that the IceCube collaboration is finalizing their $\cal{O}$(eV) sterile neutrino searches. Some preliminary results have shown significant improvement on constraining oscillation parameters~\cite{Carlos:2015,Jones:2015bya}. The IceCube bound is based on oscillation effects with a dramatical enhancement of the oscillation amplitude due to matter effects at TeV energies. Although our twin neutrino decay scenario is unlikely to be constrained by the IceCube search, it is interesting to explore how to search for the three twin neutrinos at large cosmic-ray neutrino experiments. 

In summary, motivated by the LSND and MiniBooNE excesses, we have constructed an interesting neutrino model to link the sterile neutrino phenomenology to the new TeV-scale physics associated with electroweak symmetry breaking. 

\vspace{2mm}
%%%%%%%%%%%%%%%%%%%%%%%%%%%%%%%%
\noindent
{\it{\textbf{Acknowledgments.}}}
%%%%%%%%%%%%%%%%%%%%%%%%%%%%%%%
We would like to thank Vernon Barger, Josh Berger and Lisa Everett for useful discussion and comments. This work is supported by the U. S. Department of Energy under the contract DE-FG-02-95ER40896. Jordi Salvado is also supported by FPA2011-29678 and FPA2014-57816-P.

\providecommand{\href}[2]{#2}\begingroup\raggedright\endgroup

\end{document}